\documentclass[english,aps,manuscript]{revtex4}
\usepackage[T1]{fontenc}
\usepackage[latin9]{inputenc}
\usepackage{amsmath}
\usepackage{amssymb}
\usepackage{graphicx}
\usepackage{esint}

\makeatletter
\@ifundefined{textcolor}{}
{%
 \definecolor{BLACK}{gray}{0}
 \definecolor{WHITE}{gray}{1}
 \definecolor{RED}{rgb}{1,0,0}
 \definecolor{GREEN}{rgb}{0,1,0}
 \definecolor{BLUE}{rgb}{0,0,1}
 \definecolor{CYAN}{cmyk}{1,0,0,0}
 \definecolor{MAGENTA}{cmyk}{0,1,0,0}
 \definecolor{YELLOW}{cmyk}{0,0,1,0}
}

\makeatother
\allowdisplaybreaks[3]

\makeatother

\usepackage{babel}
\begin{document}

\title{Geometric pumping induced by shear flow in dilute liquid crystalline
polymer solutions}

\author{Shunsuka Yabunaka and Hisao Hayakawa}

\affiliation{Yukawa Institute for Theoretical Physics, The Kyoto University, Kitashirakawa
Oiwake-Cho, 606-8502 Kyoto, Japan}
\begin{abstract}
We investigate nonlinear rheology of dilute liquid crystalline polymer
solutions under time dependent two-directional shear flow. We analyze
the Smoluchowski equation, which describes the dynamics of the orientation
of a liquid crystalline polymer, by employing technique of the full
counting statistics. In the adiabatic limit, we derive the expression
for time integrated currents generated by a Berry-like curvature.
Using this expression, it is shown that the expectation values of
the time-integrated angular velocity of a liquid crystalline polymer
and the time-integrated stress tensor are generally not zero even
if the time average of the shear rate is zero. The validity of the
theoretical calculations is confirmed by direct numerical simulations
of the Smoluchowski equation. Nonadiabatic effects are also investigated
by means of simulations and it is found that the time-integrated stress
tensor depends on the speed of the modulation of the shear rate if
we adopt the isotropic distribution as an initial state.
\end{abstract}
\maketitle

\section{Introduction}

Recently, nonlinear transport phenomena in stochastic processes under
an adiabatic modulation of the externally controlled parameters have
attracted much attention. It is found that a current is generated
under modulation even if there is no mean bias because of the existence
of a geometrical phase which are analogous to Berry's phase \cite{Berry,Samuel Bhandari,Thouless,Sinisyn,Sinitsyn2,Ohkubo,Chernyak}
. We call such phenomena geometric pumping in this paper. In the context
of steady state thermodynamics, Sagawa and Hayakawa obtained the geometric
expression of the excess entropy production for a Markov Jump process
\cite{Sagawa-Hayakawa}. We also note that similar phenomena in quantum
dot systems were investigated by means of quantum master equations
\cite{Ren Hanggi Li,Yuge}. The nonadiabatic effects in quantum pumping
for a spin-boson system were also investigated recently \cite{Watanabe hayakawa}.
So far, however, most of these studies were limited only to simple
jump processes and quantum dot systems. It may be of great interest
to verify whether similar effects exist even in macroscopic systems
using the framework of the geometrical pumping process. 

We focus on dilute liquid crystalline polymer solutions in this paper
as one of the simplest macroscopic systems where we can apply idea
of the geometrical pumping. Note that liquid crystalline polymers
are characterized by their high rigidity. Thus, for instance, a helix
structure of some polypeptide and tobacco mosaic viruses can be regarded
as hard rods \cite{Doi-Edwards,de Gennes Prost}. It is known that
the macroscopic rheological behavior of liquid crystalline polymer
solutions as well as the dynamics of the orientation of a liquid crystalline
polymer can be described in terms of the Smoluchowski equation \cite{Doi-Edwards}.
We analyze the Smoluchowski equation by applying the framework of
Ref. \cite{Sagawa-Hayakawa}. 

The behavior of liquid crystalline polymers under shear flows has
been studied extensively \cite{Nemoto,Ookubo}. The polymer motion
is influenced by the shear flow, while the conformation dynamics of
polymers influence the rheological properties of the polymer solution
\cite{Doi-Edwards}. Harasim et al. recently observed the dynamics
of f-actins under a steady shear flow and found that polymers undergo
tumbling motion but the periods of tumbling motion are stochastically
fluctuating \cite{Harasim}. Recently, van Leeuwen and Blöte carried
out numerical simulations of dynamics of a rod-like particle under
a shear flow in the large Weissenberg number regime and obtained asymptotic
behavior of the average of the angular velocity \cite{van Leeuwen Blote}.
We note that there are several studies on nonlinear rheology of dilute
liquid crystalline polymers, they mostly focus on steady states and
the effects of shear modulation have not been investigated systematically
\cite{Hinch Leal,Hinch Leal2,Stewart Sorensen}.

In this paper, we consider situations where there is time-dependent
two-directional shear flow. We investigate the averaged angular velocity
of rods and the macroscopic stress tensor under adiabatic modulation
of the two shear rates. It is shown that one component of the time-integrated
angular velocity of the rod and one component of the time-integrated
stress tensor are generally not zero even if the time average of the
shear rate is zero. The former result suggests that the rod can work
as a microscopic or nano machine that converts the angular velocity
of the external shear flow to the rotation of the rod under thermally
fluctuating environment. The latter result suggests that geometric
pumping effect of a single rod is related to a novel rheological response
of macroscopic dilute liquid crystalline solutions. 

The organization of this paper is as follows. In Section \ref{sec:Model},
we briefly review the Smoluchowski equation, which is employed to
describe the dynamics of orientation of a hard rod under external
flow. In Section \ref{sec:Formulation}, we formulate the full counting
statistics to derive the decomposition of the time integrated current
into the housekeeping and excess parts under adiabatic modulation
of externally controlled parameters.  In Section \ref{sec:Analysis-and-results},
we numerically calculate the excess angular velocity and the excess
stress tensor under two adiabatic shear rate protocols. We confirm
that the theoretically predicted values calculated from the expression
obtained in Section \ref{sec:Formulation} agree well with those obtained
by direct numerical simulations of the Smoluchowski equation in adiabatic
condition.

\section{Model\label{sec:Model}}

In this section, we briefly explain the model we use, and explain
the general framework of Smoluchowski equation. We consider a dilute
solution of liquid crystalline polymers under a time-dependent uniform
shear rate tensor: 
\begin{equation}
v_{\alpha}\left(\vec{x},t\right)=\kappa_{\alpha\beta}\left(t\right)x_{\beta},
\end{equation}
 where $\kappa_{\alpha\beta}\left(t\right)$ is the velocity gradient
tensor. Here, we adopt Einstein summation convention when an index
variable appears twice throughout this paper. In this paper, we consider
the two-directional shear rate tensor: 
\begin{equation}
\kappa_{\alpha\beta}\left(t\right)=\left(f_{1}\left(t\right)\delta_{\alpha x}+f_{2}\left(t\right)\delta_{\alpha y}\right)\delta_{\beta z}.\label{eq:shear rate}
\end{equation}
 We assume that the shear flow is periodic in time: $f_{i}\left(t_{f}\right)=f_{i}\left(0\right)$.
We assume that the polymers are rigid to regard them as straight hard
rods with an identical length. We denote the orientation of the long-axis
of a rod as $\vec{u}$ which satisfies $\left|\vec{u}\right|=1$.
We ignore the interactions between polymers. We also assume that the
distribution function $\Psi\left(\vec{u},t\right)$ is spatially uniform.

Each rod undergoes rotational Brownian motion as well as translational
Brownian motion due to thermal fluctuations. In overdamped cases,
the Langevin equation for the orientation of each polymer is given
by \cite{Doi-Edwards} 
\begin{equation}
\frac{d\vec{u}}{dt}=-\vec{\nabla}U\left(\vec{u}\right)+\overleftrightarrow{\kappa}\left(t\right)\cdot\vec{u}-\vec{u}\left(\vec{u}\cdot\overleftrightarrow{\kappa}\left(t\right)\cdot\vec{u}\right)+\vec{\xi},\label{eq:Rod Langevin}
\end{equation}
 where we have introduced the potential $U\left(\vec{u}\right)$ which
represents a softening constraint of $\left|\vec{u}\right|=1$ and
a zero-mean Gaussian white noise $\vec{\xi}\left(t\right)$ which
satisfies 
\begin{equation}
\left\langle \vec{\xi}\left(t\right)\vec{\xi}\left(t'\right)\right\rangle =2D_{r}\delta\left(t-t'\right)\overleftrightarrow{1},\label{eq:Gaussian white}
\end{equation}
where $D_{r}$ is the rotational diffusion constant. We note that
$\vec{u}$ is a dimensionless vector, and therefore, $D_{r}$ has
the dimension of the inverse of time. The terms $\overleftrightarrow{\kappa}\cdot\vec{u}-\vec{u}\left(\vec{u}\cdot\overleftrightarrow{\kappa}\cdot\vec{u}\right)$
represent the rotation of rods under the shear rate $\kappa_{\alpha\beta}\left(t\right)$.
It is known that Eq. (\ref{eq:Rod Langevin}) is equivalent to the
following Smoluchowski equation \cite{Doi-Edwards}.

\begin{equation}
\frac{\partial\Psi}{\partial t}=D_{r}\mathcal{R}\cdot\mathcal{R}\Psi-\mathcal{R}\cdot\left(\vec{u}\times\overleftrightarrow{\kappa}\left(t\right)\cdot\vec{u}\Psi\right)\label{eq:Smolukowski}
\end{equation}
 Here we have introduced the rotational operator $\mathcal{R}$ as
\begin{equation}
\mathcal{R}=\vec{u}\times\frac{\partial}{\partial\vec{u}}.
\end{equation}
 The first term on the right hand side (RHS) of Eq. (\ref{eq:Smolukowski})
describes the rotational diffusion of rod orientation, and the second
term represents the deterministic rotation of rods under the external
flow.

For dilute liquid crystalline polymers, the macroscopic stress tensor
$\sigma_{\alpha\beta}$ can be decomposed into two parts\cite{Doi-Edwards,Hinch Leal,Hinch Leal2}:
\begin{equation}
\sigma_{\alpha\beta}=\sigma_{\alpha\beta}^{\left(e\right)}+\sigma_{\alpha\beta}^{\left(v\right)},
\end{equation}
 where the elastic contribution $\sigma_{\alpha\beta}^{\left(e\right)}$
and the viscous contribution $\sigma_{\alpha\beta}^{\left(v\right)}$
are given by 
\begin{equation}
\sigma_{\alpha\beta}^{\left(e\right)}=k_{B}T\nu\left\langle 3\left(u_{\alpha}u_{\beta}-\frac{1}{3}\delta_{\alpha\beta}\right)\right\rangle ,
\end{equation}
 
\begin{equation}
\sigma_{\alpha\beta}^{\left(v\right)}=\frac{k_{B}T\nu}{2D_{r}}\left\langle u_{\alpha}u_{\beta}u_{\mu}u_{\nu}\right\rangle \kappa_{\mu\nu}\left(t\right),
\end{equation}
 where $\nu$ is the number of polymers in unit volume. The elastic
contribution $\sigma_{\alpha\beta}^{\left(e\right)}$ represents the
free energy increment due to the orientation alignment induced by
the external flow and the viscous contribution $\sigma_{\alpha\beta}^{\left(v\right)}$
comes from the dissipation of the free energy due to the friction
between the rod and the surrounding fluid. We note that the viscous
contribution depends on how we treat the hydrodynamic interaction
between the monomers in a liquid crystalline polymer. The coefficient
$\frac{k_{B}T\nu}{2D_{r}}$ is evaluated using the so-called shish-kebab
model \cite{Doi-Edwards,Hinch Leal,Hinch Leal2}.

Next, let us introduce the integrated angular velocity and stress
tensor $J_{1}$ and $J_{2}$: 
\begin{equation}
J_{1}=\int_{0}^{t_{f}}dt\left[\vec{u}\times\overleftrightarrow{\kappa}\left(t\right)\cdot\vec{u}\right]_{x}
\end{equation}
 
\begin{equation}
J_{2}=\int_{0}^{t_{f}}dt\sigma_{yz}=k_{B}T\nu\int_{0}^{t_{f}}dt\left[3u_{y}u_{z}+\frac{1}{2D_{r}}u_{y}u_{z}u_{\mu}u_{\nu}\kappa_{\mu\nu}\left(t\right)\right]
\end{equation}
 In this paper we consider how the statistical average of these integrated
currents $J_{1}$ and $J_{2}$ depend on the time dependent shear
rate $\kappa_{\alpha\beta}\left(t\right)$.

\section{Formulation\label{sec:Formulation}}

In this section, we consider the average value of the time integrated
quantity 
\begin{equation}
J=\int_{0}^{t_{f}}dt'H\left(\vec{u},\left\{ \kappa_{j}\left(t'\right)\right\} \right)
\end{equation}
and derive the expression of the excess part of $J$ under time dependent
shear rate $\overleftrightarrow{\kappa}\left(t\right)$. Throughout
this paper $H\left(\vec{u},\left\{ \kappa_{j}\left(t'\right)\right\} \right)$
represents an arbitrary function. To prove our results, we use the
technique of the full counting statistics \cite{Sinitsyn review,Okubo review},
which makes use of the cumulant generating function 
\begin{equation}
S\left(i\chi\right)=\ln\int dJe^{i\chi J}P\left(J\right),
\end{equation}
following Ref. \cite{Sagawa-Hayakawa} by Sagawa and Hayakawa. They
applied the adiabatic approximation to the cumulant generating function
and obtained a geometric expression of the excess currents. First,
we develop our formulation for a Brownian particle under a potential.
Next, we will impose the constraint $\left|\vec{u}\right|=1$ as by
adding a steep potential as in Eq. (\ref{eq:Rod Langevin}) and we
derive the expression of the excess part of a time-integrated quantity
for a rod.

\subsection{A Brownian particle under a potential}

We consider a Brownian particle in three dimensional space under external
force $\vec{f}\left(\vec{x},\left\{ \kappa_{j}\left(t\right)\right\} \right)$
depending on the control parameters $\left\{ \kappa_{j}\left(t\right)\right\} $.
Here, we assume that the motion of the Brownian particle is described
by the Langevin equation:

\begin{eqnarray}
\dot{\vec{x}} & = & \vec{f}\left(\vec{x},\left\{ \kappa_{j}\left(t\right)\right\} \right)+\vec{\xi}\left(t\right),\label{eq:Langevin eq}
\end{eqnarray}
 where $\vec{\xi}\left(t\right)$ is a zero-mean Gaussian white noise
satisfying 
\begin{equation}
\left\langle \vec{\xi}\left(t\right)\vec{\xi}\left(t'\right)\right\rangle =2D\delta\left(t-t'\right)\overleftrightarrow{1},\label{eq:Gaussian white-1}
\end{equation}
where $D$ is the diffusion constant.  We introduce the initial probability
distribution $P_{0}\left(\vec{x}_{0}\right)$ at the particle position
$\vec{x}_{0}$. Here, we only consider a protocol satisfying $\left\{ \kappa_{j}\left(t_{f}\right)\right\} =\left\{ \kappa_{j}\left(0\right)\right\} $.
In order to write down the path-integral expression of path probability,
we first discretize time $t$ as 
\begin{equation}
t_{n}=n\epsilon
\end{equation}
 for $n=0,1,\cdots,N$ and we take $\epsilon$ as $N\epsilon=t_{f}$.
We will take the continuous-time limit $\epsilon\rightarrow0$ in
deriving the continuous-time expression. By employing It$\bar{\mathrm{o}}$
calculus, we discretize the Langevin equation given by Eq. (\ref{eq:Langevin eq})
as 
\begin{eqnarray}
\vec{x}_{n+1} & = & \vec{x}_{n}+\vec{f}\left(\vec{x}_{n},\left\{ \kappa_{j}\left(\epsilon n\right)\right\} \right)\epsilon+\vec{\xi}_{n}\epsilon\nonumber \\
 & = & \vec{x}_{0}+\epsilon\sum_{i=0}^{n}\left(\vec{f}\left(\vec{x}_{i},\left\{ \kappa_{j}\left(\epsilon i\right)\right\} \right)+\vec{\xi}_{i}\right).\label{eq:discretized evolution}
\end{eqnarray}
 Here we now convert the Gaussian noise in Eq. (\ref{eq:Langevin eq})
by $\vec{\xi}_{i}$ $(i=0,1,\cdots,N-1)$ which satisfies 
\begin{equation}
<\vec{\xi}_{i}>=0,\quad\left\langle \vec{\xi}_{i}\vec{\xi}_{j}\right\rangle =\frac{2D}{\epsilon}\delta_{ij}\overleftrightarrow{1}.
\end{equation}
 The distribution function for $\vec{\xi}_{i}$ is given by 
\begin{equation}
P\left(\vec{\xi}_{0},\cdots,\vec{\xi}_{N-1}\right)=\left(\frac{\epsilon}{4\pi D}\right)^{3N/2}\exp\left[-\frac{\epsilon}{4D}\sum_{i=0}^{N-1}\vec{\xi}_{i}^{2}\right]
\end{equation}
 Then we convert the variable $\vec{\xi}_{i}$ to $\vec{x}_{i}$ in
the above distribution function $P\left(\vec{\xi}_{0},\cdots,\vec{\xi}_{N-1}\right)$
by using Eq. (\ref{eq:discretized evolution}). The Jacobi matrix
$\left(\frac{\partial x_{k}}{\partial\xi_{l}}\right)$ is an upper
triangular matrix if we take the row elements as $\left\{ x_{N,1},x_{N,2},x_{N,3},x_{N-1,1},x_{N-1,2},x_{N-1,3},\cdots,x_{1,1},x_{1,2},x_{1,3}\right\} $
and the line elements as $\left\{ \xi_{N-1,1},\xi_{N-1,2},\xi_{N-1,3},\xi_{N-2,1},\xi_{N-2,2},\xi_{N-2,3},\cdots,\xi_{0,1},\xi_{0,2},\xi_{0,3}\right\} $.
Then, the Jacobian of the transformation $\vec{\xi}_{i}$ to $\vec{x}_{i}$
is given by 
\begin{equation}
\left|\frac{\partial\left(\vec{x}_{1},\cdots,\vec{x}_{N}\right)}{\partial\left(\vec{\xi}_{0},\cdots,\vec{\xi}_{N-1}\right)}\right|=\epsilon^{3N}.
\end{equation}
 Thus we obtain the probability distribution function for a discretized
path $\vec{x}_{0},\vec{x}_{1},\cdots,\vec{x}_{N}$ : 
\begin{equation}
P\left(\vec{x}_{0},\vec{x}_{1},\cdots,\vec{x}_{N}\right)=\frac{1}{\epsilon^{3N}}\left(\frac{\epsilon}{4\pi D}\right)^{3N/2}\exp\left[-\frac{\epsilon}{4D}\sum_{i=0}^{N-1}\left(\frac{\vec{x}_{i+1}-\vec{x}_{i}}{\epsilon}-\vec{f}\left(\vec{x}_{i},\left\{ \kappa_{j}\left(\epsilon i\right)\right\} \right)\right)^{2}\right].
\end{equation}
 Then the cumulant generating function for this discretized process
is expressed as 
\begin{align}
S\left(i\chi\right) & =\ln\int dJe^{i\chi J}P\left(J\right)\nonumber \\
 & =\ln\int d^{3}xF_{N}\left(i\chi,\vec{x}\right),
\end{align}
 where $\chi\in\mathbb{R}$ is the counting field. Here we have introduced
the probability distribution function $P\left(J\right)$ and a function
$F_{m}\left(i\chi,\vec{x}\right)$ $(m=1,2,\cdots,N)$ as 
\begin{equation}
P\left(J\right)=\frac{1}{\epsilon^{3N}}\left(\frac{\epsilon}{4\pi D}\right)^{3N/2}\int\prod_{i=0}^{N}d^{3}x_{i}P\left(\vec{x}_{0},\vec{x}_{1},\cdots,\vec{x}_{N}\right)\delta\left(J-\epsilon\sum_{i=0}^{N-1}H\left(\vec{x}_{i},\left\{ \kappa_{j}\left(\epsilon i\right)\right\} \right)\right)P_{0}\left(\vec{x}_{0}\right),\label{eq:P(J)}
\end{equation}
\begin{eqnarray}
F_{m}\left(i\chi,\vec{x}\right) & = & \frac{1}{\epsilon^{3m}}\left(\frac{\epsilon}{4\pi D}\right)^{3m/2}\int\prod_{i=0}^{m-1}d^{3}x_{i}\exp\left[i\chi\epsilon\sum_{i=0}^{m-1}H\left(\vec{x}_{i},\left\{ \kappa_{j}\left(\epsilon i\right)\right\} \right)\right]\nonumber \\
 &  & \times\exp\left[-\frac{\epsilon}{4D}\sum_{i=0}^{m-1}\left(\frac{\vec{x}_{i+1}-\vec{x}_{i}}{\epsilon}-\vec{f}\left(\vec{x}_{i},\left\{ \kappa_{j}\left(\epsilon i\right)\right\} \right)\right)^{2}\right]P_{0}\left(\vec{x}_{0}\right),\label{eq:F_N}
\end{eqnarray}
where $\vec{x}_{m}$ is fixed as $\vec{x}_{m}=\vec{x}$ in Eq. (\ref{eq:F_N}).
We can interpret $F_{m}\left(i\chi,\vec{x}\right)$ as the path-integration
of the path probability multiplied by the phase factor $\exp\left[i\chi\int_{0}^{m\epsilon}H\left(\vec{x}\left(t\right),\left\{ \kappa_{j}\left(t'\right)\right\} \right)dt\right]\cong\exp\left[i\chi\epsilon\sum_{i=0}^{m-1}H\left(\vec{x}_{i},\left\{ \kappa_{j}\left(t'\right)\right\} \right)\right]$
with respect to the all paths which satisfy $\vec{x}\left(m\epsilon\right)=\vec{x}$.
In the continuous time limit $\epsilon\rightarrow0$, we denote the
continuous time limit of $F_{m}\left(i\chi,\vec{x}\right)$ as 
\begin{equation}
F\left(i\chi,\vec{x},t\right)\cong F_{m}\left(i\chi,\vec{x}\right),\label{eq:conlitinuous limit of F_N}
\end{equation}
where $t\cong m\epsilon$. From the definition of $F\left(i\chi,\vec{x},t\right)$
in Eqs. (\ref{eq:F_N}) and (\ref{eq:conlitinuous limit of F_N}),
we express the average value of $\int_{0}^{t}dt'H\left(\vec{x},\left\{ \kappa_{j}\left(t'\right)\right\} \right)$
as 
\begin{eqnarray}
 &  & \left\langle \int_{0}^{t}dt'H\left(\vec{x},\left\{ \kappa_{j}\left(t'\right)\right\} \right)\right\rangle \nonumber \\
 & = & \frac{1}{i}\left(\frac{\partial S\left(i\chi\right)}{\partial\chi}\right)_{\chi=0}\nonumber \\
 & = & \int d^{3}x\frac{1}{i}\left(\frac{\partial F\left(i\chi,\vec{x},t\right)}{\partial\chi}\right)_{\chi=0}\left(\int d^{3}x'F\left(0,\vec{x'},t\right)\right)^{-1}\nonumber \\
 & = & \int d^{3}x\frac{1}{i}\left(\frac{\partial F\left(i\chi,\vec{x},t\right)}{\partial\chi}\right)_{\chi=0},\label{eq:expectation of H}
\end{eqnarray}
 where $\left\langle \cdots\right\rangle $ represents the statistical
average. To obtain the last expression we have used the following
relation.
\begin{equation}
\int d^{3}x'F\left(0,\vec{x'},t\right)=1
\end{equation}

Next, let us evaluate $F\left(i\chi,\vec{x},t\right)$. As in Appendix
\ref{sec:Derivation-of-Eq.}, 
\begin{eqnarray}
\frac{dF}{dt}\left(i\chi,\vec{x},t\right) & = & -\frac{\partial}{\partial x_{k}}\left[\left(\vec{f}\left(\vec{x},\left\{ \kappa_{j}\left(t\right)\right\} \right)\right)_{k}F\left(i\chi,\vec{x},t\right)\right]+D\nabla^{2}F\left(i\chi,\vec{x},t\right)\nonumber \\
 &  & +i\chi\left(H\left(\vec{x},\left\{ \kappa_{j}\left(t\right)\right\} \right)\right)F\left(i\chi,\vec{x},t\right)\nonumber \\
 & = & \mathcal{K}\left(\chi,\left\{ \kappa_{j}\left(t\right)\right\} \right)F\left(i\chi,\vec{x},t\right),\label{eq:time evolution of F}
\end{eqnarray}
 where the operator $\mathcal{K}\left(\chi,\left\{ \kappa_{j}\left(t\right)\right\} \right)$
in the last expression of Eq. (\ref{eq:time evolution of F}) acting
on an arbitrary function $a(\vec{x})$ is given by 
\begin{eqnarray}
\mathcal{K}\left(\chi,\left\{ \kappa_{j}\left(t\right)\right\} \right)a\left(\vec{x}\right) & = & -\frac{\partial}{\partial x_{k}}\left[\left(\vec{f}\left(\vec{x},\left\{ \kappa_{j}\left(t\right)\right\} \right)\right)_{k}a\left(\vec{x}\right)\right]+D\nabla^{2}a\left(\vec{x}\right)\nonumber \\
 &  & +i\chi\left(H\left(\vec{x},\left\{ \kappa_{j}\left(t\right)\right\} \right)\right)a\left(\vec{x}\right).
\end{eqnarray}
We denote the eigenfunction of $\mathcal{K}\left(\chi,\left\{ \kappa_{j}\left(t\right)\right\} \right)$
for the $n$-th largest real part by $f_{\chi}^{n}\left(\left\{ \kappa_{j}\left(t'\right)\right\} ,\vec{x}\right)$
and its eigenvalue by $\lambda_{\chi}^{n}\left(\left\{ \kappa_{j}\left(t'\right)\right\} \right)$.
When we set $\chi=0$, Eq. (\ref{eq:time evolution of F}) reduces
to the Fokker-Planck equation for the probability distribution function
$P\left(\vec{x},t\right)=F\left(0,\vec{x},t\right)$. We assume that
the Fokker-Planck operator $\mathcal{K}\left(0,\left\{ \kappa_{j}\left(t\right)\right\} \right)$
is diagonalizable and $\lambda_{\chi=0}^{0}\left(\left\{ \kappa_{j}\left(t'\right)\right\} \right)=0$
is not degenerated. When $\left|\chi\right|$ is sufficiently small,
$\mathcal{K}\left(\chi,\left\{ \kappa_{j}\left(t\right)\right\} \right)$
can be also diagonalized and $\lambda_{\chi}^{0}\left(\left\{ \kappa_{j}\left(t'\right)\right\} \right)$
is not degenerated. In order to solve the above equation, we expand
$F\left(i\chi,\vec{x},t\right)$ in terms of the eigenfunctions as
\begin{equation}
F\left(i\chi,\vec{x},t\right)=\sum_{n}c_{n}\left(t\right)\exp\left[\int_{0}^{t}dt'\lambda_{\chi}^{n}\left(\left\{ \kappa_{j}\left(t'\right)\right\} \right)\right]f_{\chi}^{n}\left(\left\{ \kappa_{j}\left(t'\right)\right\} ,\vec{x}\right).\label{eq:expansion of F}
\end{equation}
 Using Eq. (\ref{eq:time evolution of F}), we readily obtain 
\begin{eqnarray}
\dot{c}_{0}\left(t\right) & = & -\int d^{3}x\tilde{f_{\chi}^{0}}\left(\left\{ \kappa_{j}\left(t'\right)\right\} ,\vec{x}\right)\sum_{n}c_{n}\left(t\right)\exp\left[\int_{0}^{t}dt'\left[\lambda_{\chi}^{n}\left(\left\{ \kappa_{j}\left(t'\right)\right\} \right)-\lambda_{\chi}^{0}\left(\left\{ \kappa_{j}\left(t'\right)\right\} \right)\right]\right]\nonumber \\
 &  & \times\left(\frac{\partial}{\partial\kappa_{k}}f_{\chi}^{n}\left(\left\{ \kappa_{j}\left(t'\right)\right\} ,\vec{x}\right)\right)\dot{\kappa}_{k}\left(t'\right),
\end{eqnarray}
 where we have introduced the left eigenvector $\tilde{f_{\chi}^{m}}\left(\left\{ \kappa_{j}\left(t'\right)\right\} ,\vec{x}\right)$
satisfying 
\begin{equation}
\int d^{3}x\tilde{f_{\chi}^{m}}\left(\left\{ \kappa_{j}\left(t'\right)\right\} ,\vec{x}\right)f_{\chi}^{n}\left(\left\{ \kappa_{j}\left(t'\right)\right\} ,\vec{x}\right)=\delta_{n,m},\label{eq:normalization}
\end{equation}
\begin{eqnarray}
\mathcal{K}^{\dagger}\left(\chi,\left\{ \kappa_{j}\left(t\right)\right\} \right)\tilde{f_{\chi}^{n}}\left(\left\{ \kappa_{j}\left(t'\right)\right\} ,\vec{x}\right) & = & \lambda_{\chi}^{n}\tilde{f_{\chi}^{n}}\left(\left\{ \kappa_{j}\left(t'\right)\right\} ,\vec{x}\right),
\end{eqnarray}
\begin{equation}
\left(\tilde{f_{\chi}^{0}}\left(\left\{ \kappa_{j}\left(t'\right)\right\} ,\vec{x}\right)\right)_{\chi=0}=1.\label{eq:Left eigen vector chi0}
\end{equation}
 Here the operator $\mathcal{K}^{\dagger}\left(\chi,\left\{ \kappa_{j}\left(t\right)\right\} \right)$
is the conjugate to $\mathcal{K}\left(\chi,\left\{ \kappa_{j}\left(t\right)\right\} \right)$
acting on an arbitrary function $a\left(\vec{x}\right)$ as 
\begin{align}
\mathcal{K}^{\dagger}\left(\chi,\left\{ \kappa_{j}\left(t\right)\right\} \right)a\left(\vec{x}\right)= & \left(\vec{f}\left(\vec{x},\left\{ \kappa_{j}\left(t\right)\right\} \right)\right)_{k}\frac{\partial}{\partial x_{k}}a\left(\vec{x}\right)+D\nabla^{2}a\left(\vec{x}\right)\nonumber \\
 & +i\chi\left(H\left(\vec{x},\left\{ \kappa_{j}\left(t\right)\right\} \right)\right)a\left(\vec{x}\right).
\end{align}

If we modulate the control parameter much slower than a typical relaxation
rate of the system, the contributions from $c_{n}\left(t\right)\left(n\neq0\right)$
in Eq. (\ref{eq:expansion of F}) are negligible and we obtain 
\begin{eqnarray}
F\left(i\chi,\vec{x},t\right) & \cong & c_{0}\left(i\chi\right)f_{\chi}^{0}\left(\left\{ \kappa_{j}\left(t\right)\right\} ,\vec{x}\right)\exp\left[\int_{0}^{t}dt'\lambda_{\chi}^{0}\left(\left\{ \kappa_{j}\left(t'\right)\right\} \right)\right]\nonumber \\
 &  & \times\exp\left(-\int_{0}^{t}dt'\int d^{3}x\tilde{f_{\chi}^{0}}\left(\left\{ \kappa_{j}\left(t'\right)\right\} ,\vec{x}\right)\right.\nonumber \\
 &  & \left.\times\left(\frac{\partial}{\partial\kappa_{k}}f_{\chi}^{0}\left(\left\{ \kappa_{j}\left(t'\right)\right\} ,\vec{x}\right)\right)\dot{\kappa}_{k}\left(t'\right)\right),\label{eq:F_adiabatic}
\end{eqnarray}
 where we have assumed that the real part of $\lambda_{\chi}^{n}\left(\left\{ \kappa_{j}\left(t'\right)\right\} \right)-\lambda_{\chi}^{0}\left(\left\{ \kappa_{j}\left(t'\right)\right\} \right)$
is negative for $n\neq0$. We also assume 
\begin{equation}
c_{0}\left(i\chi\right)=\int d^{3}x\tilde{f_{\chi}^{0}}\left(\left\{ \kappa_{j}\left(0\right)\right\} ,\vec{x}\right)\left(f_{\chi}^{0}\left(\left\{ \kappa_{j}\left(0\right)\right\} ,\vec{x}\right)\right)_{\chi=0},\label{eq:c0}
\end{equation}
i.e. the initial state is the stationary state $\left(f_{\chi}^{0}\left(\left\{ \kappa_{j}\left(0\right)\right\} ,\vec{x}\right)\right)_{\chi=0}$
for the initial control parameter $\left\{ \kappa_{j}\left(0\right)\right\} $.
Using Eqs. (\ref{eq:expectation of H}), (\ref{eq:normalization}),
(\ref{eq:Left eigen vector chi0}), (\ref{eq:F_adiabatic}) and (\ref{eq:c0}),
we can evaluate the average $\left\langle \int_{0}^{t_{f}}dt'H\left(\vec{x},\left\{ \kappa_{j}\left(t'\right)\right\} \right)\right\rangle $
as 
\begin{align}
 & \left\langle \int_{0}^{t_{f}}dt'H\left(\vec{x},\left\{ \kappa_{j}\left(t'\right)\right\} \right)\right\rangle \nonumber \\
= & \int d^{3}x\frac{1}{i}\left(\frac{\partial F\left(i\chi,\vec{x},t_{f}\right)}{\partial\chi}\right)_{\chi=0}\nonumber \\
= & -\frac{1}{i}\int_{0}^{t_{f}}dt'\int d^{3}x\left(\frac{\partial\tilde{f_{\chi}^{0}}}{\partial\chi}\left(\left\{ \kappa_{j}\left(t'\right)\right\} ,\vec{x}\right)\sum_{k}\left(\frac{\partial}{\partial\kappa_{k}}f_{\chi}^{0}\left(\left\{ \kappa_{j}\left(t'\right)\right\} ,\vec{x}\right)\dot{\kappa}_{k}\left(t'\right)\right)\right)_{\chi=0}\nonumber \\
 & +\int_{0}^{t_{f}}dt'H_{hk}\left(\left\{ \kappa_{j}\left(t'\right)\right\} \right)\nonumber \\
 & +\int d^{3}x\frac{1}{i}\left(\frac{\partial f_{\chi}^{0}}{\partial\chi}\left(\left\{ \kappa_{j}\left(t_{f}\right)\right\} ,\vec{x}\right)\right)_{\chi=0}\nonumber \\
 & -\int d^{3}x\frac{1}{i}\left(f_{\chi}^{0}\left(\left\{ \kappa_{j}\left(t_{f}\right)\right\} ,\vec{x}\right)\int d^{3}x'\left(\frac{\partial f_{\chi}^{0}}{\partial\chi}\left(\left\{ \kappa_{j}\left(t_{f}\right)\right\} ,\vec{x'}\right)-\frac{\partial f_{\chi}^{0}}{\partial\chi}\left(\left\{ \kappa_{j}\left(0\right)\right\} ,\vec{x'}\right)\right)\right)_{\chi=0}\nonumber \\
 & +\left(\frac{1}{i}\frac{\partial c_{0}}{\partial\chi}\left(i\chi\right)\right)_{\chi=0},\label{eq:exceptation value of H}
\end{align}
 where we have defined the housekeeping part of the current as $H_{hk}\left(\left\{ \kappa_{j}\left(t'\right)\right\} \right)\equiv\frac{1}{i}\left(\frac{\partial\lambda_{\chi}^{0}}{\partial\chi}\left(\left\{ \kappa_{j}\left(t'\right)\right\} \right)\right)_{\chi=0}$.
The housekeeping part represents the average current $\left\langle H\left(\vec{x},\left\{ \kappa_{j}\left(t'\right)\right\} \right)\right\rangle $
under the steady state corresponding to the control parameters $\left\{ \kappa_{j}\left(t'\right)\right\} .$
We note the following relation 
\begin{align}
\left(\frac{\partial c_{0}\left(i\chi\right)}{\partial\chi}\right)_{\chi=0} & =\int d^{3}x\frac{\partial\tilde{f_{\chi}^{0}}}{\partial\chi}\left(\left\{ \kappa_{j}\left(0\right)\right\} ,\vec{x}\right)\left(f_{\chi}^{0}\left(\left\{ \kappa_{j}\left(0\right)\right\} ,\vec{x}\right)\right)_{\chi=0}\nonumber \\
 & =-\int d^{3}x\frac{\partial f_{\chi}^{0}}{\partial\chi}\left(\left\{ \kappa_{j}\left(0\right)\right\} ,\vec{x}\right)\left(\tilde{f_{\chi}^{0}}\left(\left\{ \kappa_{j}\left(0\right)\right\} ,\vec{x}\right)\right)_{\chi=0}\nonumber \\
 & =-\int d^{3}x\frac{\partial f_{\chi}^{0}}{\partial\chi}\left(\left\{ \kappa_{j}\left(0\right)\right\} ,\vec{x}\right),
\end{align}
where we have used Eqs. (\ref{eq:normalization}) and (\ref{eq:Left eigen vector chi0}).
From the above relation and $\left\{ \kappa_{j}\left(t_{f}\right)\right\} =\left\{ \kappa_{j}\left(0\right)\right\} $,
we finally obtain a simple expression of the statistical average as
\begin{align}
 & \left\langle \int_{0}^{t_{f}}dt'H\left(\vec{x},\left\{ \kappa_{j}\left(t'\right)\right\} \right)\right\rangle \nonumber \\
= & -\frac{1}{i}\int_{0}^{t_{f}}dt'\int d^{3}x\left(\frac{\partial\tilde{f_{\chi}^{0}}}{\partial\chi}\left(\left\{ \kappa_{j}\left(t'\right)\right\} ,\vec{x}\right)\right)_{\chi=0}\left(\frac{\partial}{\partial\kappa_{k}}f^{0}\left(\left\{ \kappa_{j}\left(t'\right)\right\} ,\vec{x}\right)\right)_{\chi=0}\dot{\kappa}_{k}\left(t'\right)\nonumber \\
 & +\int_{0}^{t_{f}}dt'H_{hk}\left(\left\{ \kappa_{j}\left(t'\right)\right\} \right),\label{eq:exceptation value of H-1}
\end{align}
Thus, the statistical average of the time-integrated quantity $\left\langle \int_{0}^{t_{f}}dt'H\left(\vec{x},\left\{ \kappa_{j}\left(t'\right)\right\} \right)\right\rangle $
under adiabatic modulation of the externally controlled parameter,
in general, can be decomposed into the excess part and the house keeping
part.

\subsection{Geometrical phase of a rigid rod under shear}

In this section, we apply the general framework developed in the last
subsection to Eq. (\ref{eq:Rod Langevin}), which describes the dynamics
of the orientation of a rod under a shear flow. We take the shear
rate $f_{1}$ and $f_{2}$ defined in Eq. (\ref{eq:shear rate}) as
external controlled parameters $\left\{ \kappa_{j}\left(t\right)\right\} $
in this case and we derive the expression of the statistical average
$\left\langle \int_{0}^{t_{f}}dt'H\left(\vec{u},\overleftrightarrow{\kappa}\left(t'\right)\right)\right\rangle $.
In the next section, we apply this result to 
\begin{equation}
H_{1}\equiv\left[\vec{u}\times\overleftrightarrow{\kappa}\left(t\right)\cdot\vec{u}\right]_{x}
\end{equation}
 and 
\begin{equation}
H_{2}\equiv\left[3u_{y}u_{z}+\frac{1}{2D_{r}}u_{y}u_{z}u_{\mu}u_{\nu}\kappa_{\mu\nu}\left(t\right)\right].
\end{equation}

In this subsection, to simplify the argument, we adopt the following
potential $U\left(\vec{u}\right)$ for the restriction $\left|\vec{u}\right|=1$,
\begin{align}
U\left(\vec{u}\right) & =0\,\left(1-\delta\leq\left|\vec{u}\right|\leq1+\delta\right)\\
 & =\infty\,\left(\mathrm{otherwise}\right)
\end{align}
 and we will take the limit $\delta\rightarrow0$, though our formulation
is also applicable to arbitrary potentials that realize the restriction
$\left|\vec{u}\right|=1$. In this case, $\left|\vec{u}\right|$ is
confined in a thin region between $1-\delta\leq\left|\vec{u}\right|\leq1+\delta$.
Let us introduce 
\begin{equation}
F'\left(i\chi,\vec{u},t\right)=\int_{1-\delta}^{1+\delta}d\alpha F\left(i\chi,\alpha\vec{u},t\right)
\end{equation}
 on the spherical surface $\left|\vec{u}\right|=1$. In the limit
$\delta\rightarrow0$, we expect that the variation of $F\left(i\chi,\alpha\vec{u},t\right)$
along the direction parallel to $\vec{u}$ is small and it can be
expressed as 
\begin{equation}
F\left(i\chi,\alpha\vec{u},t\right)\cong\frac{1}{2\delta}F'\left(i\chi,\vec{u},t\right)
\end{equation}
 By setting the diffusion constant in Eq. (\ref{eq:Gaussian white-1})
as $D=D_{r}$ with the aid of Eq. (\ref{eq:time evolution of F}),
the time evolution equation for $F'\left(i\chi,\vec{u},t\right)$
can be rewritten as 
\begin{eqnarray}
\frac{dF'}{dt}\left(i\chi,\vec{u},t\right) & \cong & -\frac{\partial}{\partial u_{j}}\left[\left(\left(\overleftrightarrow{\kappa}\cdot\vec{u}\right)_{j}-u_{j}\left(\vec{u}\cdot\overleftrightarrow{\kappa}\left(t\right)\cdot\vec{u}\Psi\right)\right)F'\left(i\chi,\vec{u},t\right)\right]+D_{r}\nabla_{\perp}^{2}F'\left(i\chi,\vec{u},t\right)\nonumber \\
 &  & +i\chi\left(H\left(\vec{u},\left\{ \overleftrightarrow{\kappa}\left(t\right)\right\} \right)\right)F'\left(i\chi,\vec{u},t\right)\nonumber \\
 & = & D_{r}\mathcal{R}\cdot\left(\mathcal{R}F'\left(i\chi,\vec{u},t\right)\right)-\mathcal{R}\cdot\left(\vec{u}\times\overleftrightarrow{\kappa}\left(t\right)\cdot\vec{u}F'\left(i\chi,\vec{u},t\right)\right)\nonumber \\
 &  & +i\chi H\left(\vec{u},\left\{ \overleftrightarrow{\kappa}\left(t\right)\right\} \right)F'\left(i\chi,\vec{u},t\right)\nonumber \\
 & = & \left(\mathcal{K}_{0}'\left(\left\{ \overleftrightarrow{\kappa}\left(t\right)\right\} \right)+i\chi H\left(\vec{u},\left\{ \overleftrightarrow{\kappa}\left(t\right)\right\} \right)\right)F'\left(i\chi,\vec{u},t\right),\label{eq:time ev of generating function}
\end{eqnarray}
 where $\nabla_{\perp}^{2}$ is the spherical Laplacian and the operator
$\mathcal{K}_{0}'\left(\left\{ \overleftrightarrow{\kappa}\left(t\right)\right\} \right)$
on the spherical surface $\left|\vec{u}\right|=1$ acting on an arbitrary
function $f\left(\vec{u}\right)$ has been defined as 
\begin{equation}
\mathcal{K}_{0}'\left(\left\{ \overleftrightarrow{\kappa}\left(t\right)\right\} \right)f\left(\vec{u}\right)=D_{r}\mathcal{R}\cdot\left(\mathcal{R}f\left(\vec{u}\right)\right)-\mathcal{R}\cdot\left(\vec{u}\times\overleftrightarrow{\kappa}\left(t\right)\cdot\vec{u}f\left(\vec{u}\right)\right)
\end{equation}
 The statistical average $\left\langle \int_{0}^{t_{f}}dt'H\left(\vec{u},\overleftrightarrow{\kappa}\left(t'\right)\right)\right\rangle $
in this case can be also derived using Eq. (\ref{eq:exceptation value of H-1})
as 
\begin{align}
 & \left\langle \int_{0}^{t_{f}}dt'H\left(\vec{u},\overleftrightarrow{\kappa}\left(t'\right)\right)\right\rangle \nonumber \\
= & -\frac{1}{i}\int_{0}^{t_{f}}dt'\int d\Omega\left(\frac{\partial\tilde{f_{\chi}^{0}}}{\partial\chi}\left(\overleftrightarrow{\kappa}\left(t'\right),\vec{u}\right)\right)_{\chi=0}\left(\frac{\partial}{\partial f_{j}}f_{\chi}^{0}\left(\overleftrightarrow{\kappa}\left(t'\right),\vec{u}\right)\right)_{\chi=0}\dot{f}_{j}\left(t'\right)\nonumber \\
 & +\int_{0}^{t_{f}}dt'H_{hk}\left(\overleftrightarrow{\kappa}\left(t'\right)\right),\label{eq:exceptation value of H-1-1}
\end{align}
where $f_{\chi}^{0}\left(\overleftrightarrow{\kappa}\left(t'\right),\vec{u}\right)$
and $\tilde{f_{\chi}^{0}}\left(\overleftrightarrow{\kappa}\left(t'\right),\vec{u}\right)$
are, respectively, the right and left eigenfunction of the operator
$\mathcal{K}_{0}'\left(\left\{ \overleftrightarrow{\kappa}\left(t\right)\right\} \right)+i\chi H$
corresponding to the eigenvalue $\lambda_{\chi}^{0}\left(\overleftrightarrow{\kappa}\left(t\right)\right)$.
$\int d\Omega$ denotes integration over $\left|\vec{u}\right|=1$.
The right eigenfunction $f_{\chi}^{0}\left(\overleftrightarrow{\kappa}\left(t'\right),\vec{u}\right)$,
now, satisfies the eigenequation 
\begin{align}
 & D_{r}\mathcal{R}\cdot\left(\mathcal{R}f_{\chi}^{0}\left(\overleftrightarrow{\kappa}\left(t'\right),\vec{u}\right)\right)-\mathcal{R}\cdot\left(\left(\vec{u}\times\overleftrightarrow{\kappa}\left(t'\right)\cdot\vec{u}\right)f_{\chi}^{0}\left(\overleftrightarrow{\kappa}\left(t'\right),\vec{u}\right)\right)\nonumber \\
 & +i\chi H\left(\vec{u},\left\{ \overleftrightarrow{\kappa}\left(t\right)\right\} \right)f_{\chi}^{0}\left(\overleftrightarrow{\kappa}\left(t'\right),\vec{u}\right)\nonumber \\
= & \lambda_{\chi}^{0}\left(\overleftrightarrow{\kappa}\left(t'\right)\right)f_{\chi}^{0}\left(\overleftrightarrow{\kappa}\left(t'\right),\vec{u}\right).\label{eq:right eigen equation}
\end{align}
Using Green's theorem, the first term on RHS in Eq. (\ref{eq:exceptation value of H-1-1})
can be rewritten as 
\begin{equation}
\mp\frac{1}{i}\int_{C}dS\int d\Omega\epsilon_{3jk}\frac{\partial}{\partial f_{k}}\left(\frac{\partial\tilde{f_{\chi}^{0}}}{\partial\chi}\left(\overleftrightarrow{\kappa}\left(t'\right),\vec{u}\right)\right)_{\chi=0}\frac{\partial}{\partial f_{j}}\left(f_{\chi}^{0}\left(\overleftrightarrow{\kappa}\left(t'\right),\vec{u}\right)\right)_{\chi=0},
\end{equation}
where $\epsilon_{3jk}$ is the Levi-Civita symbol in three dimensional
space and $C$ is the domain in $f_{1}-f_{2}$ space which is surrounded
by the closed path $f_{1}\left(t\right)$ and $f_{2}\left(t\right)$
introduced in Eq. (\ref{eq:shear rate}) for $0\le t\le t_{f}$. We
take the $-$ ($+$) sign if the path $(f_{1}\left(t\right),f_{2}\left(t\right))$
circulates around the domain $C$ clockwise (anti-clockwise). Here
we can regard 
\begin{equation}
-\frac{1}{i}\int d\Omega\epsilon_{3jk}\frac{\partial}{\partial f_{k}}\left(\frac{\partial\tilde{f_{\chi}^{0}}}{\partial\chi}\left(\overleftrightarrow{\kappa}\left(t'\right),\vec{u}\right)\right)_{\chi=0}\frac{\partial}{\partial f_{j}}\left(f_{\chi}^{0}\left(\overleftrightarrow{\kappa}\left(t'\right),\vec{u}\right)\right)_{\chi=0}\label{eq:Berry-like curvature}
\end{equation}
 as a Berry-like curvature in the parameter space which creates the
excess time-integrated current by an adiabatic modulation of the shear
rates.

\section{Analysis and results\label{sec:Analysis-and-results}}

\begin{figure}
\includegraphics[scale=0.35]{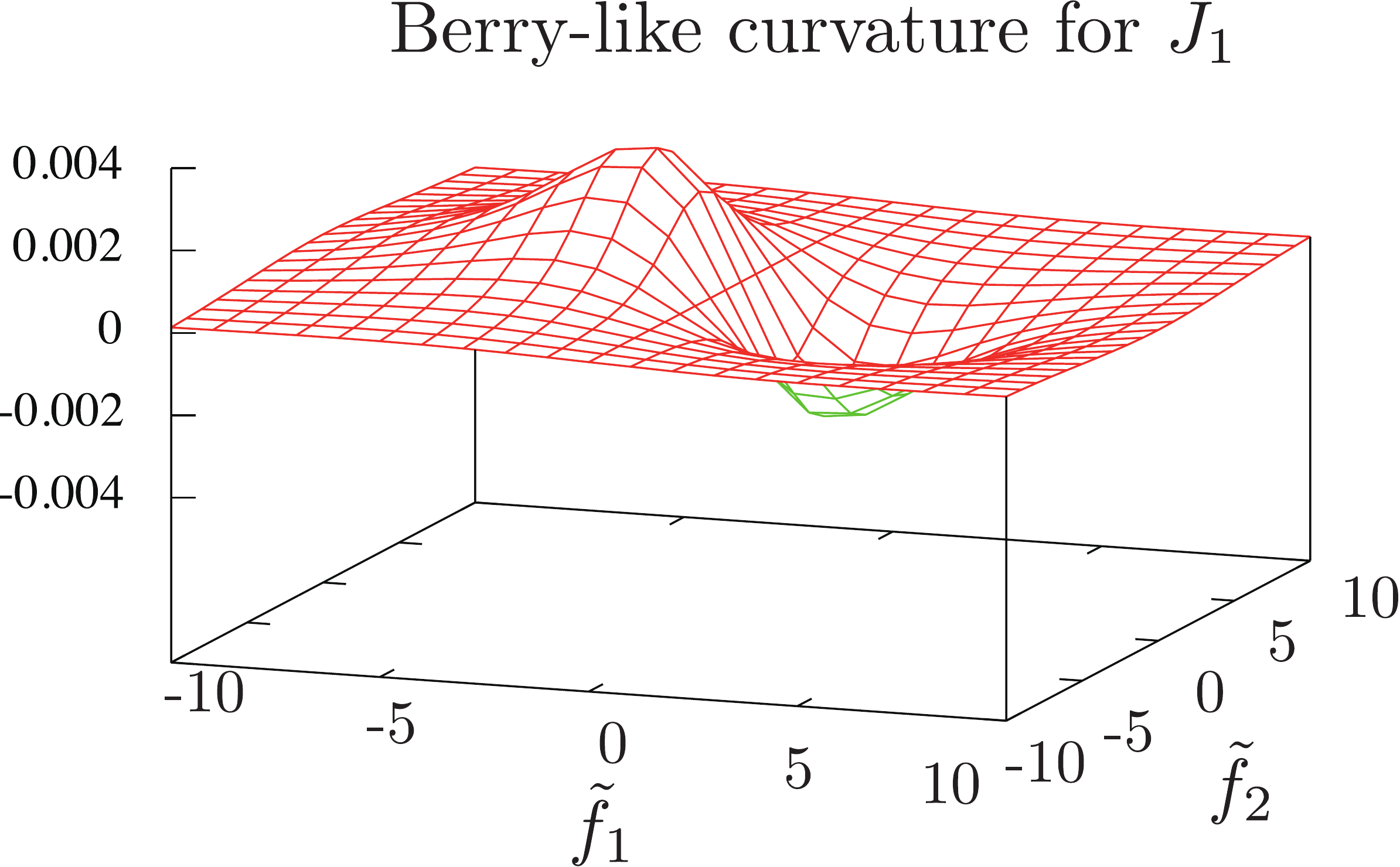}\includegraphics[scale=0.35]{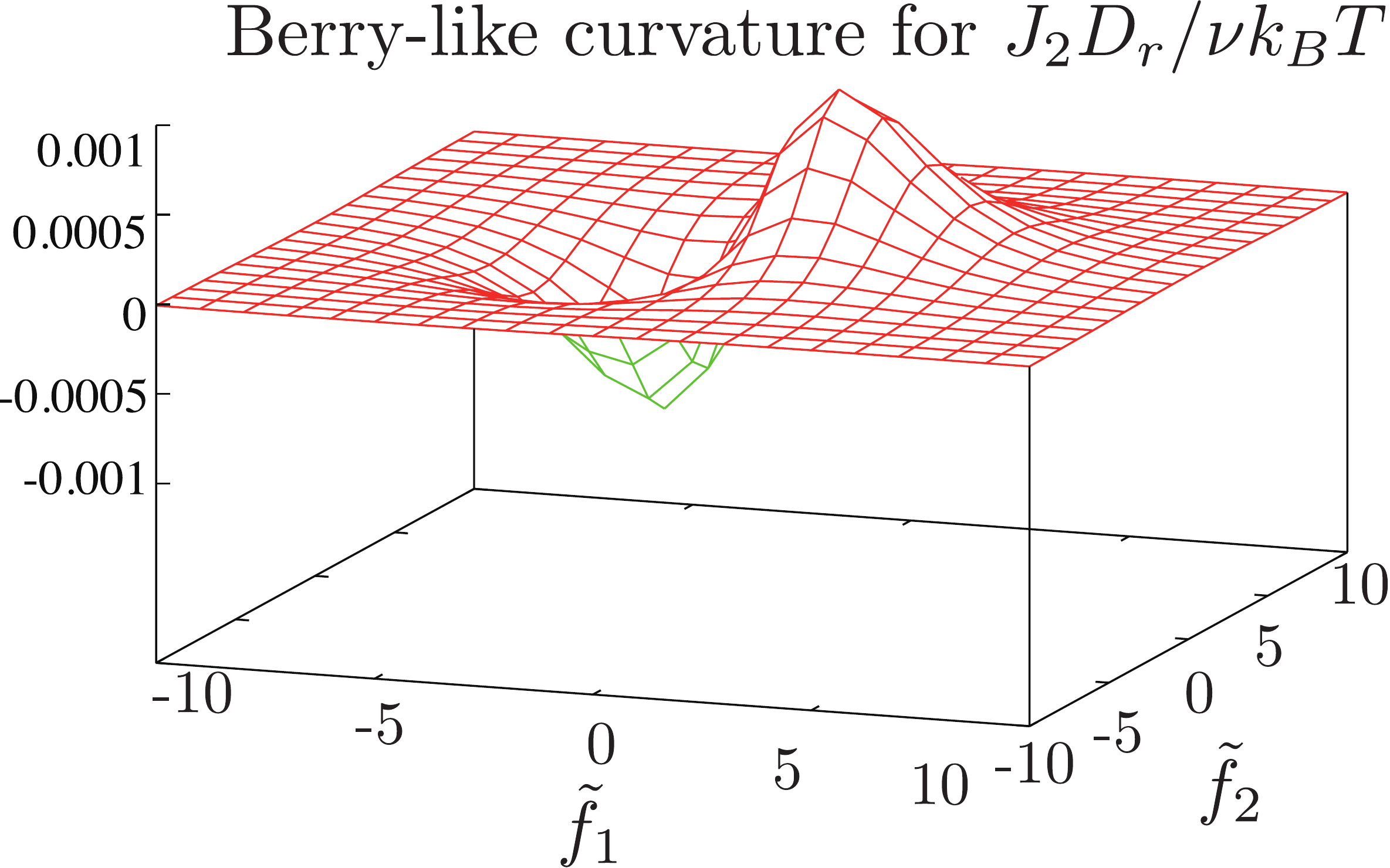}\caption{The Berry-like curvatures in $\tilde{f_{1}}-\tilde{f_{2}}$ plane
associated with the time-integrated angular velocity $J_{1}$ (Left)
and the scaled time-integrated stress tensor $J_{2}D_{r}/\nu k_{B}T$
(Right).}
\label{Flo:curvature rotation}
\end{figure}

Now let us calculate the excess currents $J_{1}$ and $J_{2}$ using
Eq. (\ref{eq:exceptation value of H-1-1}). Several formulae used
in the following calculations are given in Appendix \ref{sec:Several-formulae-used}.
We need to find the eigenvalue $\lambda_{\chi}^{0}\left(\left\{ \overleftrightarrow{\kappa}\left(t\right)\right\} \right)$
and its corresponding right and left eigenvectors $f_{\chi}^{0}\left(\overleftrightarrow{\kappa},\vec{u}\right)$
and $\tilde{f_{\chi}^{0}}\left(\overleftrightarrow{\kappa},\vec{u}\right)$
for the operator $\mathcal{K}_{0}'\left(\left\{ \overleftrightarrow{\kappa}\left(t\right)\right\} \right)+i\chi H_{i}$.
For this purpose, let us expand $f_{\chi}^{0}\left(\overleftrightarrow{\kappa},\vec{u}\right)$
and $\tilde{f_{\chi}^{0}}\left(\overleftrightarrow{\kappa},\vec{u}\right)$
in terms of the spherical harmonics $Y_{lm}\left(\vec{u}\right)$
as

\begin{equation}
f_{\chi}^{0}\left(\overleftrightarrow{\kappa},\vec{u}\right)=\sum_{l=0}^{\infty}\sum_{m=-l}^{l}a_{lm}\left(\chi,\overleftrightarrow{\kappa}\right)Y_{lm}\left(\vec{u}\right).
\end{equation}
 For $l=1,2,\cdots$ and $m=-l,-l+1,\cdots,l-1,l$, the spherical
harmonics $Y_{lm}\left(\vec{u}\right)$ are given by Eq. (\ref{eq:Y_lm}).
Using Eqs. (\ref{eq:exceptation value of H-1-1}), (\ref{eq:matrix K0})
and (\ref{eq:matrix H1}), we can construct a finite-dimensional matrix
by projecting the operator $\mathcal{K}_{0}'\left(\left\{ \overleftrightarrow{\kappa}\left(t\right)\right\} \right)+i\chi H_{i}$
to a space spanned by finite number of spherical harmonics. By solving
its eigenequation Eq. (\ref{eq:right eigen equation}), we can evaluate
the Berry-like curvatures Eq. (\ref{eq:Berry-like curvature}) for
the excess part of the time-integrated currents $J_{1}$ and $J_{2}$.
Here we truncate the spherical harmonic functions up to $l=10$ and
numerically evaluate the Berry-like curvatures as presented in Fig.
\ref{Flo:curvature rotation}, where we use the scaled shear rate
$f_{i}=D_{r}\tilde{f_{i}}$. Note that, in the cases of steady shear,
the dimensionless number $\sqrt{\tilde{f}_{1}^{2}+\tilde{f}_{2}^{2}}$
is known as Weissenberg number. 

To check the validity of the theoretical calculations, we simulate
Langevin equation Eq. (\ref{eq:Rod Langevin}) under the following
two shear protocols, where the housekeeping contributions in Eq. (\ref{eq:exceptation value of H-1-1})
vanish in adiabatic condition. For $0\leq t\leq t_{f}$, protocol
I:

\begin{equation}
\left(\tilde{f_{1}}\left(t\right),\tilde{f_{2}}\left(t\right)\right)=\left(5-5\cos\left(2\pi t/t_{f}\right),5\sin\left(2\pi t/t_{f}\right)\right),
\end{equation}
and protocol II: 
\begin{equation}
\left(\tilde{f_{1}}\left(t\right),\tilde{f_{2}}\left(t\right)\right)=\begin{cases}
\left(0,30t/t_{f}\right) & \left(t<\frac{1}{6}t_{f}\right)\\
\left(30\left(t/t_{f}-\frac{1}{6}\right),5\right) & \left(\frac{1}{6}t_{f}\leq t<\frac{2}{6}t_{f}\right)\\
\left(5,30\left(\frac{1}{2}-t/t_{f}\right)\right) & \left(\frac{2}{6}t_{f}\leq t<\frac{4}{6}t_{f}\right)\\
\left(30\left(\frac{5}{6}-t/t_{f}\right),-5\right) & \left(\frac{4}{6}t_{f}\leq t<\frac{5}{6}t_{f}\right)\\
\left(0,30\left(t/t_{f}-1\right)\right) & \left(\frac{5}{6}t_{f}\leq t<t_{f}\right)
\end{cases}.
\end{equation}
To evaluate the dependence of $J_{1}$ and $J_{2}$ on the initial
distribution of $\vec{u}$, we repeat each protocol twice, namely,
\begin{equation}
\left(\tilde{f_{1}}\left(t\right),\tilde{f_{2}}\left(t\right)\right)=\left(\tilde{f_{1}}\left(t-t_{f\mathbb{}}\right),\tilde{f_{2}}\left(t-t_{f\mathbb{}}\right)\right)\quad\left(t_{f}\leq t\leq2t_{f}\right).
\end{equation}
This choice implies that the measurement starts from $F'\left(0,\vec{u},t_{f}\right)$
in the second cycle. In numerical simulations, we discretize the time
with $\Delta t=0.002D_{r}^{-1}$ and we investigate how the time integrated
currents $J_{1}$ and $J_{2}$ obtained from simulations approaches
to the theoretically predicted values when we increase $t_{f}$. As
the initial distribution of $\vec{u}$ at $t=0$, we adopt the isotropic
distribution, which is the steady state for $\left(f_{1}(0),f_{2}(0)\right)=(0,0)$.
Then, we expect that $J_{1}\left(J_{2}\right)$ integrated during
$0\leq t\leq t_{f}$ is equal to $J_{1}\left(J_{2}\right)$ integrated
during $t_{f}\leq t\leq2t_{f}$ in the adiabatic limit $t_{f}\rightarrow\infty$.
We take statistical average of the time integrated quantities $J_{1}$
and $J_{2}$ over $N=100000\sim8000000$ ensembles, depending on $t_{f}$,
for the protocols I and II. We note that larger number of ensembles
is needed for larger $t_{f}$ to reduce the statistical error. The
error bars of the numerical results are evaluated by calculating the
dispersions of the sample averages $\sigma_{a}$ using the relation
\begin{equation}
\sigma_{a}^{2}=\frac{\sigma^{2}}{N-1},
\end{equation}
where $\sigma$ is the dispersion of the sample itself. We also evaluate
the theoretical value by numerically integrating the Berry-like curvature
Eq. (\ref{eq:Berry-like curvature}) inside the domains $C_{I}$ and
$C_{II}$ surrounded by the path $(f_{1}\left(t\right),f_{2}\left(t\right))\,\left(0\leq t\leq t_{f}\right)$
in protocol I and II, respectively. We note that $(f_{1}\left(t\right),f_{2}\left(t\right))\,\left(0\leq t\leq t_{f}\right)$
circulates clockwise around $C_{I(II)}$ in protocol I(II). 

\begin{figure}

\includegraphics[scale=0.27]{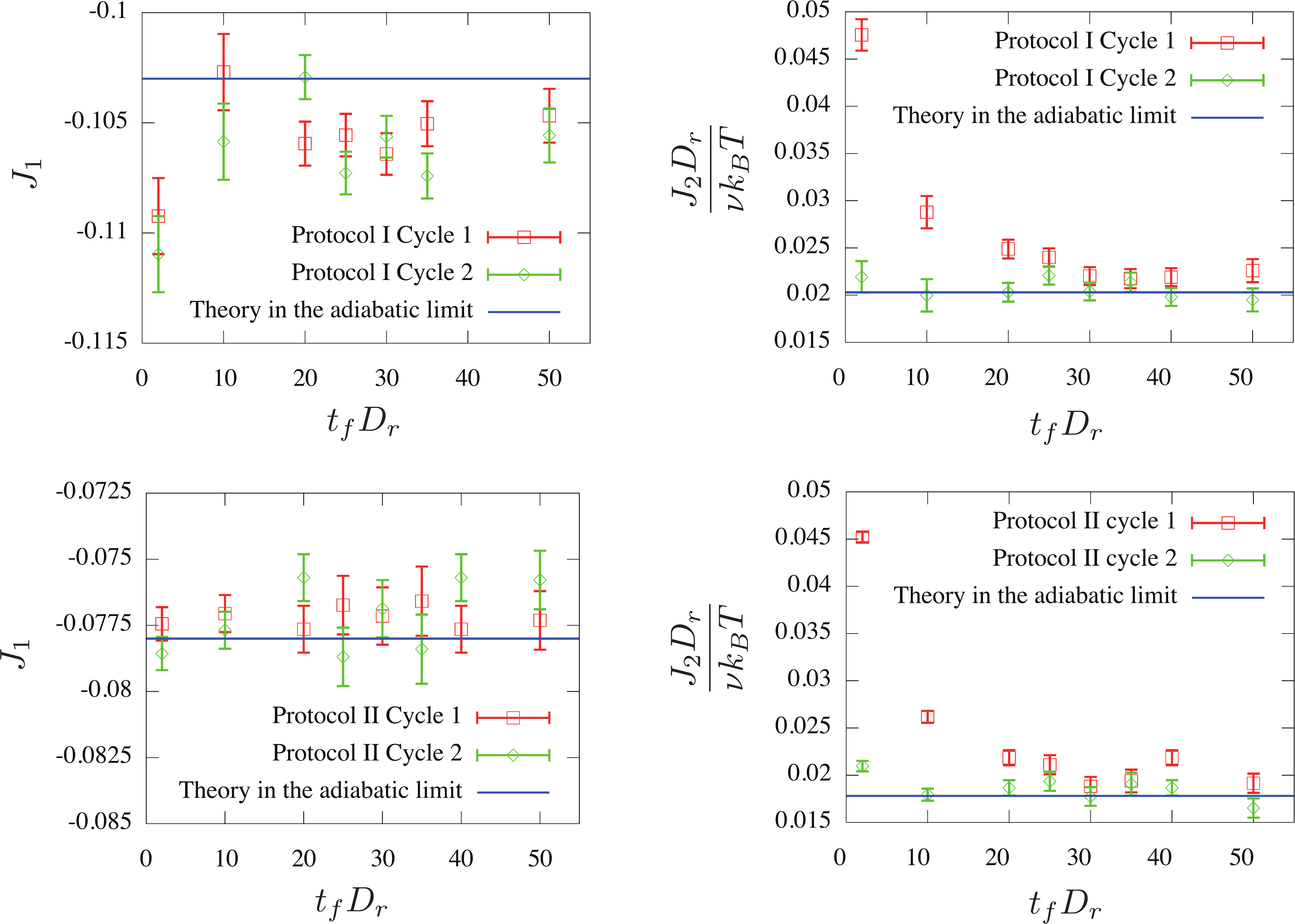}

\caption{Comparison between the theoretically predicted values and the results
of numerical simulations.}
\label{comparison}

\end{figure}

The comparison between the theoretical values and the simulated values
is shown in Fig. \ref{comparison}. Here we plot the simulated value
of $J_{1}$ and $J_{2}$ integrated during the first cycle $\left(0\leq t\leq t_{f}\right)$
and the second cycle $\left(t_{f}\leq t\leq2t_{f}\right)$ to compare
them with the theoretical values predicted in the adiabatic limit.
As for $J_{1}$, the values obtained from both the first cycle and
the second cycle except for $t_{f}=2$ agree with the theoretical
values within the sum of the statistical errors (shown by the error
bars) and the systematic errors, which are of the order of a few percent
of the theoretical values. The systematic error may be due to the
discrete time interval used in our simulations. As for $J_{2}$, the
values obtained from the first cycle are much larger for $t_{f}\lesssim25D_{r}^{-1}$
than the theoretical values, which suggests $J_{2}$ for smaller $t_{f}$
has large dependence on the initial condition. We can also see that
the values of $J_{2}$ obtained from the second cycle agree well with
the theoretical values within the error bars. 


\section{Discussions}

In this paper, we have analyzed the Smoluchowski equation under a
time-dependent shear. We have found that the time-integrated angular
velocity and the time-integrated stress tensor are generated due to
the geometrical phase such as Berry-like phase created by an adiabatic
modulation of the shear rate even when the housekeeping contributions
vanish. It has been shown that the expectation values of the time-integrated
angular velocity of a liquid crystalline polymer and the time-integrated
stress tensor are generally not zero even if the time average of the
shear rate is zero. The validity of our theoretical calculation has
been verified by the direct simulation of the Smoluchowski equation.
We have confirmed that the theoretically predicted values calculated
from the expression obtained in Section \ref{sec:Formulation} agree
well with those obtained by direct numerical simulations of the Smoluchowski
equation in adiabatic condition ($t_{f}\rightarrow\infty$).

We can interpret that such excess contributions originates from the
deviation of the orientation distribution function from the steady
state distribution function as follows. When we modulate the external
parameter slowly compared with the rotational diffusion, $\left(\dot{\overleftrightarrow{\kappa}}\left(t\right)\right)$
is proportional to the slowness parameter $\epsilon$. We expect that
$F'\left(0,\vec{u},t\right)$ can be expanded in terms of $\epsilon$
as 
\begin{equation}
F'\left(0,\vec{u},t\right)\cong\left(f_{\chi}^{0}\left(\overleftrightarrow{\kappa}\left(t\right),\vec{u}\right)\right)_{\chi=0}+\epsilon\delta F'\left(0,\vec{u},t\right)+O\left(\epsilon^{2}\right),
\end{equation}
where $\delta F'\left(0,\vec{u},t\right)$ is of the order $O\left(1\right)$.
Therefore, $\epsilon\delta F'\left(0,\vec{u},t\right)$ gives the
$O\left(1\right)$ correction to the time-integrated current $J_{i}$
because the required time for a protocol is the order of $O\left(\epsilon^{-1}\right)$.

We examined the dependence of the integrated currents $J_{1}$ and
$J_{2}$ on the initial condition.  Our results suggest that the
dependence of $J_{1}$ on the initial condition is not so strong,
which is contrast to the result for a spin-boson system \cite{Watanabe hayakawa}.
The values of $J_{2}$ obtained from the numerical simulations are
much larger than the theoretically predicted values if we take the
isotropic distribution as an initial condition for $t_{f}\lesssim25D_{r}^{-1}$,
while $J_{2}$ obtained from the numerical simulation agrees well
with the theoretical prediction if we adopt the distribution $F'\left(0,\vec{u},t_{f}\right)$
obtained numerically as an initial condition. This effect should be
regarded as an nonadiabatic effect \cite{Watanabe hayakawa} because
$F'\left(0,\vec{u},t_{f}\right)$ approaches to the isotropic distribution
in the limit $t_{f}\rightarrow\infty$. 

We note that our formulation in Section \ref{sec:Formulation} for
Langevin systems is quite general and applicable to other systems.
For example, the motion of a Brownian particle under an adiabatically
modulated trapping potential can be discussed within our formulation.

$J_{1}$ can be measured by means of direct observation of the orientation
of a rod under shear. However it requires large number of ensembles
(about 100000 ensembles even for $t_{f}=2D_{r}^{-1}$ ) in order to
reduce the statistical error. $J_{2}$ can be measured by macroscopic
rheological experiments on a dilute liquid crystalline solution. In
such cases, we expect that it is not needed to make statistical average
because there exist large number of liquid crystalline polymers in
the solution.

We have discussed the rheology of dilute liquid crystalline solutions,
which can be described only by the one-body distribution function
of the orientation vectors $\vec{u}$ of rods. It is of interest to
investigate whether path-dependent rheological response also exists
in other fluid systems where the interactions between particles are
important and spatial correlations exist. To investigate such systems,
we need to extend our formulation for Liouville equations which describe
interacting particle systems or fluctuating hydrodynamic equations.

\section{Acknowledgement}

One of the authors (S.Y.) was supported by the Japan Society for Promotion
of Science. We are grateful to Yusuke Korai for useful discussions
and letting us know the formula Eq. (\ref{eq:integral of the product of 3 spherical harmonics}).
This work is partially supported by the Grant-in-Aid of MEXT (Grant
No. 25287098).

\appendix

\section{Derivation of Eq. (\ref{eq:time evolution of F}) \label{sec:Derivation-of-Eq.}}

In this section we derive Eq. (\ref{eq:time evolution of F}). At
$t=m\epsilon$, we approximate $\frac{dF}{dt}\left(i\chi,\vec{x},t\right)$
as $\frac{dF}{dt}\left(i\chi,\vec{x},t\right)\cong\frac{1}{\epsilon}\left(F_{m}\left(i\chi,\vec{x}\right)-F_{m-1}\left(i\chi,\vec{x}\right)\right)$.
Then it can be evaluated as
\begin{eqnarray}
\frac{dF}{dt}\left(i\chi,\vec{x},t\right) & \cong & \frac{1}{\epsilon}\left(F_{m}\left(i\chi,\vec{x}\right)-F_{m-1}\left(i\chi,\vec{x}\right)\right)\nonumber \\
 & = & \frac{1}{\epsilon}\left(\frac{1}{\epsilon^{3}}\left(\frac{\epsilon}{4\pi D}\right)^{3/2}\int d^{3}x_{m-1}\exp\left[i\chi\epsilon\left(H\left(\vec{x}_{m-1},\left\{ \kappa_{j}\left(\epsilon\left(m-1\right)\right)\right\} \right)\right)\right]\right.\nonumber \\
 &  & \left.\times\exp\left[-\frac{\epsilon}{4D}\left(\frac{\vec{x}-\vec{x}_{m-1}}{\epsilon}-\vec{f}\left(\vec{x}_{m-1},\left\{ \kappa_{j}\left(\epsilon\left(m-1\right)\right)\right\} \right)\right)^{2}\right]F_{m-1}\left(i\chi,\vec{x}_{m-1}\right)\right)\nonumber \\
 &  & -\frac{1}{\epsilon}F_{m-1}\left(i\chi,\vec{x}\right).
\end{eqnarray}
Here we have used the following relation, which can be derived from
the path-integral expression of $F_{m-1}\left(i\chi,\vec{x}_{m-1}\right)$
given by Eq. (\ref{eq:F_N}). 
\begin{eqnarray}
F_{m}\left(i\chi,\vec{x}\right) & = & \frac{1}{\epsilon^{3}}\left(\frac{\epsilon}{4\pi D}\right)^{3/2}\int d^{3}x_{m-1}\exp\left[i\chi\epsilon\left(H\left(\vec{x}_{m-1},\left\{ \kappa_{j}\left(\epsilon\left(m-1\right)\right)\right\} \right)\right)\right]\nonumber \\
 &  & \times\exp\left[-\frac{\epsilon}{4D}\left(\frac{\vec{x}-\vec{x}_{m-1}}{\epsilon}-\vec{f}\left(\vec{x}_{m-1},\left\{ \kappa_{j}\left(\epsilon\left(m-1\right)\right)\right\} \right)\right)^{2}\right]\nonumber \\
 &  & \times F_{m-1}\left(i\chi,\vec{x}_{m-1}\right)
\end{eqnarray}
By expanding $F_{m-1}\left(i\chi,\vec{x}_{m-1}\right)$ around $\vec{x}$:
\begin{align}
F_{m-1}\left(i\chi,\vec{x}_{m-1}\right)= & F_{m-1}\left(i\chi,\vec{x}\right)-\left(\vec{x}-\vec{x}_{m-1}\right)_{k}\frac{\partial}{\partial x_{k}}F_{m-1}\left(i\chi,\vec{x}\right)\nonumber \\
 & +\frac{1}{2}\left(\vec{x}-\vec{x}_{m-1}\right)_{k}\left(\vec{x}-\vec{x}_{m-1}\right)_{l}\frac{\partial^{2}}{\partial x_{k}\partial x_{l}}F_{m-1}\left(i\chi,\vec{x}\right)+\cdots,
\end{align}
we can derive the following equation in the continuous time limit
$\left(\epsilon\rightarrow0\right)$ by keeping only the terms $O\left(1\right)$.
\begin{eqnarray}
\frac{dF}{dt}\left(i\chi,\vec{x},t\right) & \cong & \frac{1}{\epsilon}\left(\frac{1}{\epsilon^{3}}\left(\frac{\epsilon}{4\pi D}\right)^{3/2}\int d^{3}x_{m-1}\exp\left[-\frac{\epsilon}{4D}\left(\frac{\vec{x}-\vec{x}_{m-1}}{\epsilon}-\vec{f}\left(\vec{x}_{m-1},\left\{ \kappa_{j}\left(\epsilon\left(m-1\right)\right)\right\} \right)\right)^{2}\right]\right.\nonumber \\
 &  & \times\left[1+i\chi\epsilon H\left(\vec{x}_{m-1},\left\{ \kappa_{j}\left(\epsilon\left(m-1\right)\right)\right\} \right)+O\left(\epsilon^{2}\right)\right]\nonumber \\
 &  & \times\left(F_{m-1}\left(i\chi,\vec{x}\right)-\left(\vec{x}-\vec{x}_{m-1}\right)_{k}\frac{\partial}{\partial x_{k}}F_{m-1}\left(i\chi,\vec{x}\right)\right.\nonumber \\
 &  & \left.+\frac{1}{2}\left(\vec{x}-\vec{x}_{m-1}\right)_{k}\left(\vec{x}-\vec{x}_{m-1}\right)_{l}\frac{\partial^{2}}{\partial x_{k}\partial x_{l}}F_{m-1}\left(i\chi,\vec{x}\right)+\cdots\right)\nonumber \\
 &  & -\frac{1}{\epsilon}F_{m-1}\left(i\chi,\vec{x}\right)\nonumber \\
 & = & -\left(\vec{f}\left(\vec{x},\left\{ \kappa_{j}\left(\epsilon\left(m-1\right)\right)\right\} \right)\right)_{k}\frac{\partial}{\partial x_{k}}F_{m-1}\left(i\chi,\vec{x}\right)+D\nabla^{2}F_{m-1}\left(i\chi,\vec{x}\right)\nonumber \\
 &  & -\left(\frac{\partial}{\partial x_{k}}\left(\vec{f}\left(\vec{x},\left\{ \kappa_{j}\left(\epsilon\left(m-1\right)\right)\right\} \right)\right)_{k}\right)F_{m-1}\left(i\chi,\vec{x}\right)\nonumber \\
 &  & +i\chi H\left(\vec{x},\left\{ \kappa_{j}\left(\epsilon\left(m-1\right)\right)\right\} \right)F_{m-1}\left(i\chi,\vec{x}\right)
\end{eqnarray}
Here we have employed the following equations. 
\begin{align}
 & \frac{1}{\epsilon}\left(\frac{1}{\epsilon^{3}}\left(\frac{\epsilon}{4\pi D}\right)^{3/2}\int d^{3}x_{m-1}\exp\left[-\frac{\epsilon}{4D}\left(\frac{\vec{x}-\vec{x}_{m-1}}{\epsilon}-\vec{f}\left(\vec{x}_{m-1},\left\{ \kappa_{j}\left(\epsilon\left(m-1\right)\right)\right\} \right)\right)^{2}\right]\right.\nonumber \\
= & \frac{1}{\epsilon}-\sum_{k}\frac{\partial}{\partial x_{k}}\left(\vec{f}\left(\vec{x},\left\{ \kappa_{j}\left(\epsilon\left(m-1\right)\right)\right\} \right)\right)_{k}+O\left(\epsilon\right),
\end{align}
 
\begin{align}
 & \frac{1}{\epsilon}\frac{1}{\epsilon^{3}}\left(\frac{\epsilon}{4\pi D}\right)^{3/2}\int d^{3}x_{m-1}\exp\left[-\frac{\epsilon}{4D}\left(\frac{\vec{x}-\vec{x}_{m-1}}{\epsilon}-\vec{f}\left(\vec{x}_{m-1},\left\{ \kappa_{j}\left(\epsilon\left(m-1\right)\right)\right\} \right)\right)^{2}\right]\left(\vec{x}-\vec{x}_{m-1}\right)_{k}\nonumber \\
= & \left(\vec{f}\left(\vec{x},\left\{ \kappa_{j}\left(\epsilon\left(m-1\right)\right)\right\} \right)\right)_{k}+O\left(\epsilon\right),
\end{align}
 
\begin{align}
 & \frac{1}{\epsilon}\frac{1}{\epsilon^{3}}\left(\frac{\epsilon}{4\pi D}\right)^{3/2}\int d^{3}x_{m-1}\exp\left[-\frac{\epsilon}{4D}\left(\frac{\vec{x}-\vec{x}_{m-1}}{\epsilon}-\vec{f}\left(\vec{x}_{m-1},\left\{ \kappa_{j}\left(\epsilon\left(m-1\right)\right)\right\} \right)\right)^{2}\right]\nonumber \\
 & \times\left(\vec{x}-\vec{x}_{m-1}\right)_{k}\left(\vec{x}-\vec{x}_{m-1}\right)_{l}\nonumber \\
= & 2D\delta_{kl}+O\left(\epsilon\right).
\end{align}

\section{Several formulae used in the evaluation of the berry-like curvatures\label{sec:Several-formulae-used}}

In this section, we give several formulae used in the calculations
in Section \ref{sec:Analysis-and-results}. For $l=1,2,\cdots$ and
$m=-l,-l+1,\cdots,l-1,l$ , the spherical harmonics is defined as 

\begin{equation}
Y_{lm}\left(\vec{u}\right)=\left(-1\right)^{\left(m+\left|m\right|\right)/2}\sqrt{\frac{2l+1}{4\pi}}\sqrt{\frac{\left(l-\left|m\right|\right)!}{\left(l+\left|m\right|\right)!}}P_{l}^{\left|m\right|}\left(\cos\theta\right)e^{im\phi},\label{eq:Y_lm}
\end{equation}
 where $P_{l}^{\left|m\right|}$ are associated Legendre polynomials.
We note that the polar coordinates $\theta$ and $\phi$ are connected
to $\vec{u}$ as 
\begin{equation}
\vec{u}=\left(\cos\theta\cos\phi,\cos\theta\sin\phi,\sin\phi\right).
\end{equation}
 The matrix representations of $\mathcal{K}_{0}'$ and $H_{i}$ for
$i=1,2$ are given by 
\begin{align}
 & \left\langle lm\left|\mathcal{K}_{0}'\right|l'm'\right\rangle \nonumber \\
= & -l\left(l+1\right)\delta_{ll'}\delta_{mm'}\nonumber \\
 & -\frac{1}{3}\sqrt{\frac{\left(2l'+1\right)}{\left(2l+1\right)}}\left\langle l,0\left|2,0,\right.l'0\right\rangle \times\nonumber \\
 & \left[\left(f_{1}-if_{2}\right)\sqrt{l\left(l+1\right)-\left(m-1\right)m}\left\langle l,\left(m-1\right)\left|2,0,\right.l'm'\right\rangle \right.\nonumber \\
 & \left.+\left(-f_{1}-if_{2}\right)\sqrt{l\left(l+1\right)-\left(m+1\right)m}\left\langle l,\left(m+1\right)\left|2,0,\right.l'm'\right\rangle \right]\nonumber \\
 & +\left(-f_{1}+if_{2}\right)\frac{1}{6}\sqrt{l\left(l+1\right)-\left(m-1\right)m}\delta_{m-1,m'}\delta_{l,l'}\nonumber \\
 & +\left(f_{1}+if_{2}\right)\frac{1}{6}\sqrt{l\left(l+1\right)-\left(m+1\right)m}\delta_{m+1,m'}\delta_{l,l'}\nonumber \\
 & +m\left\langle l,0\left|2,0,\right.l'0\right\rangle \sqrt{\frac{\left(2l'+1\right)}{6\left(2l+1\right)}}\nonumber \\
 & \times\left(\left(-f_{1}+if_{2}\right)\left\langle l,m\left|2,1,\right.l'm'\right\rangle +\left(-f_{1}-if_{2}\right)\left\langle l,m\left|2,-1,\right.l'm'\right\rangle \right),\label{eq:matrix K0}
\end{align}
 
\begin{equation}
\left\langle lm\left|H_{1}\right|l'm'\right\rangle =-f_{2}\frac{1}{3}\left(2\sqrt{\frac{\left(2l'+1\right)}{\left(2l+1\right)}}\left\langle l,0\left|2,0,\right.l'0\right\rangle \left\langle l,m\left|2,0,\right.l'm'\right\rangle +\delta_{ll'}\delta_{mm'}\right),\label{eq:matrix H1}
\end{equation}
\begin{align}
 & \left\langle lm\left|H_{2}\right|l'm'\right\rangle \nonumber \\
= & \sqrt{\frac{\left(2l'+1\right)}{\left(2l+1\right)}}\times\nonumber \\
 & \left[\left\langle l,0\left|4,0,\right.l'0\right\rangle \times\left(-\frac{1}{7\sqrt{10}}\left(if_{1}+f_{2}\right)\left\langle l,m\left|4,2,\right.l',m'\right\rangle \right.\right.\nonumber \\
 & \left.+\frac{1}{7\sqrt{10}}\left(if_{1}-f_{2}\right)\left\langle l,m\left|4,-2,\right.l',m'\right\rangle -\frac{2}{35}f_{1}\left\langle l,m\left|4,0,\right.l',m'\right\rangle \right)\nonumber \\
 & +\left\langle l,0\left|2,0,\right.l'0\right\rangle \times\nonumber \\
 & \left(\frac{1}{14}\sqrt{\frac{1}{6}}\left(-if_{1}-f_{2}\right)\left\langle l,m\left|2,2,\right.l',m'\right\rangle +\sqrt{\frac{3}{2}}i\left\langle l,m\left|2,1,\right.l',m'\right\rangle \right.\nonumber \\
 & +\frac{1}{42}f_{2}\left\langle l,m\left|2,0,\right.l',m'\right\rangle +\sqrt{\frac{3}{2}}i\left\langle l,m\left|2,-1,\right.l',m'\right\rangle \nonumber \\
 & \left.+\frac{1}{14}\sqrt{\frac{1}{6}}\left(if_{1}-f_{2}\right)\left\langle l,m\left|2,-2,\right.l',m'\right\rangle \right)\nonumber \\
 & \left.+\frac{1}{30}f_{2}\delta_{ll'}\delta_{mm'}\right].\label{eq:matrix H2}
\end{align}
Here we have employed the Dirac's bra-ket notation and $\left|l,m\right\rangle $
represents $Y_{lm}\left(\vec{u}\right)$. We have also introduced
the Clebsch-Gordan coefficients $\left\langle l,m\left|l'm',\right.l''m''\right\rangle $.
In deriving these expressions, we have used the following formula
\cite{Messiah}.
\begin{align}
 & \int d\Omega\left(Y_{lm}\left(\vec{u}\right)\right)^{*}Y_{l_{1}m_{1}}\left(\vec{u}\right)Y_{l_{2}m_{2}}\left(\vec{u}\right)\nonumber \\
= & \sqrt{\frac{\left(2l_{1}+1\right)\left(2l_{2}+1\right)}{4\pi\left(2l+1\right)}}\left\langle l0\left|l_{1}0,\right.l_{2}0\right\rangle \left\langle lm\left|l_{1}m_{1},\right.l_{2}m_{2}\right\rangle \label{eq:integral of the product of 3 spherical harmonics}
\end{align}

\end{document}